\documentclass[a4paper]{ISOStyle}
\usepackage{epsfig}

\newcommand{\micron}{\mbox{$\mu$m}}%

\begin{document}

\title{Far-Infrared ISO Limits on Dust Disks around Millisecond
        Pulsars}

\author{T.\,J.\,W.\,Lazio\inst{1} \and J.\,Fischer\inst{1} \and
	R.\,S.\,Foster\inst{1,2}}
  
  \institute{Naval Research Laboratory, Code~7213, Remote Sensing Division, 
         Washington, DC 20375-5351 USA
  \and 
     Present address: Booz-Allen \& Hamilton Inc., 8283~Greensboro
	Drive, McLean, VA  22102-3838 USA
}

\maketitle 

\begin{abstract}
We report 60 and~90~\micron\ observations of 7 millisecond pulsars
with the \hbox{ISOPHOT} instrument and describe our analysis
procedures.  No pulsars were detected, and typical ($3\sigma$) upper
limits are 150~mJy.  We combine our results with others in the
literature and use them to place constraints on the existence of
protoplanetary or dust disks around millisecond pulsars.

\keywords{Accretion, accretion disks -- planetary systems:
	protoplanetary disks -- pulsars: general }
\end{abstract}

\section{INTRODUCTION}\label{LFF_sec:intro}

The first extrasolar planets discovered were found around the
millisecond pulsar \object{PSR~B1257$+$12} (\cite{wf92}).  Various
mechanisms have been proposed for their formation (\cite{ph92}), but
all generally rely on an accretion disk around the pulsar within which
the planets form.  Millisecond pulsars themselves are understood to be
neutron stars that have been ``spun-up'' by the accretion of matter
from a companion (\cite{v95}).

Both of these arguments suggest that dust disks may exist around
millisecond pulsars.  These dust disks may exist regardless of whether
or not the pulsar is orbited by planets.  The planetary formation
process is unlikely to be 100\% efficient, and a dust disk may
represent the remnant debris from the formation of the planetary system.
Alternately, not all accretion disks may give rise to planetary
systems, and a dusk disk would represent the remnant of the initial
accretion that spun-up the pulsar.

\begin{figure}[bht]
  \begin{center}
    \epsfig{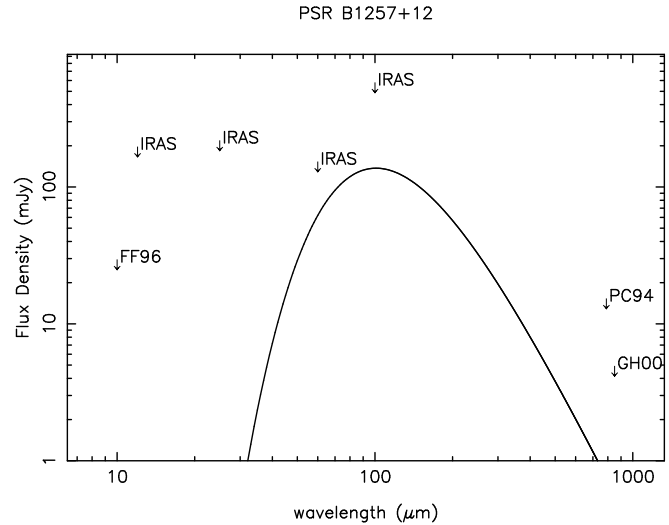}
  \end{center}
\caption{Upper limits on the infrared and sub-millimeter emission from
\object{PSR~B1257$+$12}.  Limits are from Foster \& Fischer~(1996,
FF96 and IRAS), Phillips \& Chandler~(1994, PC94), and Greaves \&
Holland~(2000, GH00).  The solid line shows the expected emission from 
a 300~M${}_{\oplus}$ dust disk composed of dust particles 0.1~\micron\ 
in size and heated by 1\% of the spin-down luminosity of the pulsar
(Foster \& Fischer~1996).  The pulsar is assumed to be at a distance of~620~pc.}
\label{LFF_fig:b1257+12}
\end{figure}

A modest number of unsuccessful searches for infrared emission from
dust around millisecond pulsars have been conducted.
Figure~\ref{LFF_fig:b1257+12} compares the current observational
limits for \object{PSR~B1257$+$12} with the predicted emission levels
from a dust disk model.  The limits for other pulsars are similar.
Figure~\ref{LFF_fig:b1257+12} illustrates the need for far-infrared
observations.

\section{OBSERVATIONS}\label{LFF_sec:observe}

In order to generate a millisecond pulsar ISOPHOT target list, we
compiled a list of millisecond pulsars known prior to 1994~August and
with distances less than 1~kpc.  Distances are estimated from the
\cite{tc93} model and should be accurate to approximately 25\%.  Most
of these millisecond pulsars lie at high Galactic latitudes.  

Of these, seven were observed with the ISOPHOT instrument
(\cite{lemkeetal96}) onboard the ISO satellite (\cite{kessleretal96}).
Table~\ref{LFF_tab:log} summarizes the observing details.  All of the
observations used the P32 observing mode with the C100 detector.  In
this mode the spacecraft was commanded to cover a series of raster
pointings around the nominal pulsar position.  At each raster pointing
an internal chopper pointed the beam toward~13 adjacent sky positions.
The throw of the chopper was larger than the offset between raster
pointings.  The result was that, in general, an individual sky
position within the raster was observed multiple times or oversampled.
Before and after each observation of a pulsar, an internal calibration
source was observed.

\begin{table}[bht]
\caption{Pulsars Observed}\label{LFF_tab:log}
\begin{center}
 \leavevmode
 \footnotesize
 \begin{tabular}{lccccc}
  \hline \\[-5pt]
         &             &        &                     & {P32}    & {} \\
  {Name} & {$\lambda$} & {$D$}  & {$L_{\mathrm{sd}}$} & {Raster} & {Time} \\
  {PSR}  & {(\micron)} & {(pc)} & {(L${}_\odot$)}     &          & {(s)} \\[+5pt]
      \hline \\[-5pt]

\object{J0034$-$0534} & 90 & 1000 & 10   & 3 $\times$ 8 & 1402 \\

\object{J1640$+$2224} & 60 & 1190 & 0.88 & 3 $\times$ 6 & 1012 \\
                      & 90 & 1190 & 0.88 & 3 $\times$ 8 & 848 \\

\object{J1730$-$2304} & 60 &  510 & $<$ 0.35 & 3 $\times$ 6 & 1590 \\
	              & 90 &  510 & $<$ 0.35 & 3 $\times$ 8 & 1232 \\

\object{B1855$+$09}   & 60 &  900 & 1.1      & 3 $\times$ 6 & 1012 \\

\object{J2124$-$3358} & 60 &  240 & 1.7      & 3 $\times$ 6 & 1012 \\
	              & 90 &  240 & 1.7      & 3 $\times$ 8 & 848 \\

\\

\object{J2145$-$0750} & 60 &  500 & $<$ 0.048 & 3 $\times$ 6 & 1590 \\
	              & 90 &  500 & $<$ 0.048 & 3 $\times$ 8 & 848 \\

\object{J2322$+$2057} & 60 &  780 & 0.62     & 3 $\times$ 6 & 1804 \\
	              & 90 &  780 & 0.62     & 3 $\times$ 8 & 1402 \\

  \hline\\

 \end{tabular}
\end{center}
\end{table}

The analysis of the pulsar observations largely followed the standard
ISOPHOT analysis.  The key difference was the amount of
``deglitching'' performed.  Glitches result from cosmic rays striking
the detector or secondary electrons produced by spacecraft materials
struck by primary cosmic rays.  Failure to remove glitches can corrupt
the responsivity drift correction of \emph{all} data, not just those
containing the glitches themselves.  The standard ISOPHOT analysis
pipeline removes glitches but does so without making use of the
redundancy implicit in the oversampled P32 observations.

Deglitching proceeded in the following fashion.  Within each
spacecraft pointing the chopper would sweep past a particular sky
position multiple times (typically 3--5 times).  For each sky
position, the median signal level was determined, then subtracted from
all observations at that sky position.  The observations from all sky
positions were then combined to form a signal strength histogram.  A
signal strength threshold was specified, and signals above this level
were eliminated.  Typically 3\%--10\% of the signals were eliminated
in this stage.  Depending upon the number of chopper sweeps per
spacecraft pointing and deglitching prior to this stage, the median
signal strength per sky position could not always be determined
accurately.  In these cases, additional manual deglitching was done to
remove any remaining outlier signals.  Our use of the observations of
the internal calibration sources followed the standard ISOPHOT
analysis pipeline.

After deglitching and calibration using the internal calibration
sources, mapping was done within the ISOPHOT Interactive Analysis
package.  Measurements from the individual detector pixels were
co-added to form a sky image, with the contributions from the individual 
detector pixels weighted by their distances from the image pixels.
Doing so takes into account the beam profile falling on each detector
pixel.  We also employed a median flat field, which has the effect of
reducing substantially our sensitivity to any extended emission in the 
field but increasing our sensitivity to point sources, as desired for
this program.

\begin{figure}[bht]
  \begin{center}
    \epsfig{file=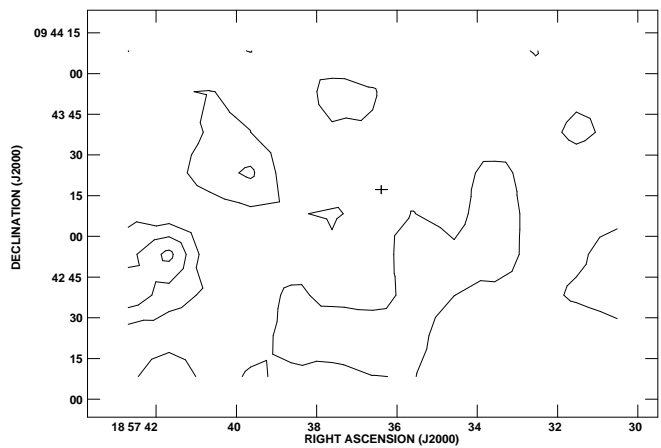, width=6.5cm, angle=-90}
  \end{center}
\caption{\object{PSR~B1855$+$09} at~60~\micron.  The rms noise level
is 50~mJy, and the contour levels are 1.5, 1.6, 1.7, and~1.8~Jy.  The 
cross marks the location of the pulsar, and the size of the cross is
5000 times larger than the uncertainty in the pulsar's location and
proper motion.}
\label{LFF_fig:b1855+09}
\end{figure}

As an example of the resulting images produced,
Figure~\ref{LFF_fig:b1855+09} shows the image of
\object{PSR~B1855$+$09}.  In no case have we identified a
point source at the location of a pulsar.  Utilizing the inner quarter
of the image, we determined the rms noise level.  We take our upper
limits to be 3 times this rms noise level.  Typical (1$\sigma$) values
are 50~mJy, with higher values seen for pulsars close to the Galactic
plane.

\section{RESULTS}\label{LFF_sec:results}

Foster \& Fischer~(1996) developed a model for the infrared emission
from a dust disk around a millisecond pulsar.  Their model assumes
that the disk consists of particles of a uniform radius~$a$ heated by
a fraction~$f_{\mathrm{sd}}$ of the pulsar's spin-down
luminosity~L${}_{\mathrm{sd}}$.  The total mass of the disk is $m_d$.
While the model is simplistic---an actual dust disk presumably
consists of particles with a range of sizes, the heating mechanism is
left unspecified, and non-equilibrium effects such as stochastic
heating are ignored---we believe that this simplicity is justified
given the uncertainties of the heating mechanism and of the environs
of a millisecond pulsar.

For $f_{\mathrm{sd}} \sim 1$\%, typical dust temperatures are
predicted to be $T \approx 10$--50~K for disks having $m_d \sim
100$~M${}_\oplus$ and~$a \sim 1$~\micron\ and illuminated by a pulsar
with L${}_{\mathrm{sd}} \sim 1$~L${}_\odot$.  These temperatures are
considerably lower than those assumed ($T \approx 150$~K) by Phillips
\& Chandler~(1994), who estimated disk temperatures by scaling from
observations of T~Tauri stars.  The lower temperatures result from our
assumption of a weaker coupling between the pulsar's spin-down
luminosity and the disk.  Phillips \& Chandler~(1994) considered disk
temperature to be a major uncertainty in converting from measured flux
densities to inferred disk masses.  Accordingly, our assumption of a
weaker coupling means that larger disk masses can be tolerated without
violating the observational constraints.

In addition to these observations with ISO, other infrared and
sub-millimeter observations of millisecond pulsars have been conducted.
Those observations most relevant to our sample of millisecond pulsars
are those by Foster \& Fischer~(1996) at~10~\micron\ and Greaves \&
Holland~(2000) at~850~\micron.  Unfortunately, there is little overlap
between these three samples of pulsars.  Most of the pulsars
that have been observed between~10 and~850~\micron\ have been observed
at only one or two wavelengths.  In general, it is therefore not
possible to constrain all three parameters of this model with the
existing observations.

\begin{figure}[bht]
  \begin{center}
    \epsfig{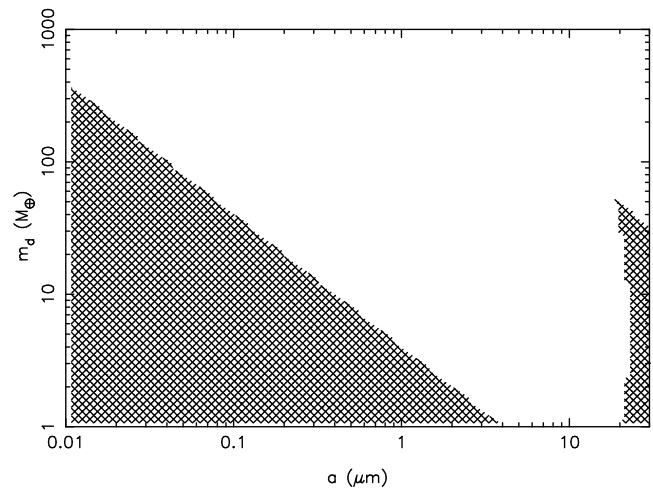}
  \end{center}
\caption{Allowed disk masses and grain sizes for a dust disk orbiting
\object{PSR~J0034$-$0534}.  The cross-hatched region indicates disk
masses and grain sizes that do not violate the observational limits
at~90 and~850~\micron.  A heating efficiency of $f_{\mathrm{sd}} = 
1$\% has been assumed.}
\label{LFF_fig:like2}
\end{figure}

We therefore adopt an approach in which we infer limits on two
parameters of the Foster \& Fischer~(1996) model for fiducial values
of the third parameter.  Elsewhere we shall consider the detailed
implications for the well-studied objects from our sample
(\cite{lff01}).  Here, as an example, we consider the millisecond
pulsar \object{PSR~J0034$-$0534} (\cite{bailesetal94}) which has a
probable white dwarf companion, is at a distance of~1~kpc, and has a
spin-down luminosity of~10~L${}_\odot$.  Greaves \& Holland~(2000)
placed a $2\sigma$ limit of~3.7~mJy at 850~\micron, and we place a
$2\sigma$ limit of~50~mJy at~90~\micron.  Figure~\ref{LFF_fig:like2}
shows the allowed region of the disk mass-grain size plane given these
observational limits and an assumed heating efficiency of
$f_{\mathrm{sd}} = 1$\%.

Allowed regions in the $m_d$-$a$ plane occur for one of two possible
reasons.  First, the peak of the dust disk emission may appear
shortward of~90~\micron, where no constraints exist for this pulsar,
with the Rayleigh-Jeans tail of the emission falling below the two
measured values.  This region is to the lower left in
Figure~\ref{LFF_fig:like2}.  Second, the peak of the emission may
appear between~90~\micron\ and~850~\micron, but with a magnitude
comparable to that measured at~90~\micron\ so that the Rayleigh-Jeans
tail again does not violate the 850~\micron\ limit while the Wien tail
of the emission does not violate the 90~\micron\ limit.  This region
is to the lower right in Figure~\ref{LFF_fig:like2}.  Obviously, a
lower value of~$f_{\mathrm{sd}}$ would produce larger allowed regions
in the $m_d$-$a$ plane.

\section{CONCLUSIONS}

We have reported 60 and~90~\micron\ observations of 7 millisecond
pulsars with the \hbox{ISOPHOT} instrument.  We have described our
analysis procedures, which utilized the standard ISOPHOT Interactive
Analysis package but also relied on considerably more ``deglitching''
than is usually the case.  No pulsars were detected, and typical
($3\sigma$) upper limits are 150~mJy.

The current set of pulsars for which mid-infrared, far-infrared, and
sub-millimeter observations exists is dominated by pulsars having
measurements at only 1 or~2 wavelengths.  Measurements at additional
wavelengths are required in order to use existing dust disk emission
models to place meaningful constraints on the presence or absence of a 
circumpulsar dust disk.

\begin{acknowledgements}

We thank the organizers of the ISOPHOT Workshop on PHT32 Oversampled
Mapping, particularly R.~Tuffs, C.~Gabriel, N.~Lu, and B.~Schulz for
their many helpful discussions, and R.~Tuffs for his deglitching
software.  Without their assistance, no results would be reported
here.  The results reported here are based on observations with ISO,
an ESA project with instruments funded by ESA Member States
(especially the PI countries: France, Germany, the Netherlands and the
United Kingdom) and with the participation of ISAS and NASA.  The
ISOPHOT data presented in this paper were reduced using PIA, which is
a joint development by the ESA Astrophysics Division and the ISOPHOT
consortium, with the collaboration of the Infrared Analysis and
Processing Center (IPAC) and the Instituto de Astrof{\'\i}sica de
Canarias (IAC).  This work was supported partially by the NASA ISO
grant program.  Basic research in astronomy at the NRL is supported by
the Office of Naval Research.

\end{acknowledgements}

\end{document}